\documentstyle[aps,psfig,prl]{revtex}
\begin{document}
\draft
\twocolumn
\title{Fluctuations of the Condensate in Ideal and Interacting Bose Gases} 

\author{ Hongwei Xiong$^{1,2}$, Shujuan Liu$^{1}$, Guoxiang Huang$^{3}$, Zhijun Xu$^{1}$, Cunyuan Zhang$^{1}$}

\address{$^{1}$Department of Applied Physics, Zhejiang 
University of Technology, Hangzhou, 310032, P. R. China} 
\address{$^{2}$Zhijiang College,  
Zhejiang University of Technology, Hangzhou, 310012, P. R. China }
\address{$^{3}$Department of Physics, East China Normal University, 200062, Shanghai, P. R. China}

\date{\today}

\maketitle

\begin{abstract}

{\it We investigate the fluctuations of the condensate in the ideal and weakly interacting
Bose gases confined in a box of volume $V$ within canonical ensemble.
Canonical ensemble is developed to describe the behavior of the
fluctuations when different methods of approximation to the weakly interacting
Bose gases are used. Research shows
 that the fluctuations of the condensate exhibit anomalous
behavior for the interacting Bose gas confined in a box.
} 

\end{abstract}


\narrowtext

\section{Introduction}

The experimental achievement of Bose-Einstein condensation (BEC) in dilute alkali
atoms \cite{ALK}, spin-polarized hydrogen \cite{MIT}
and recently in metastable helium \cite{HEL}
has enormously
stimulated the theoretical research \cite{RMP}
on the ultracold bosons. In particular, fluctuations
$\left\langle \delta ^{2}N_{\bf{0}}\right\rangle $ of the mean ground state occupation number
$N_{\bf{0}}$
have been recently thoroughly investigated in a series of 
papers.
Apart from the intrinsic theoretical interest, it is foreseeable that such fluctuations
will become experimentally testable in the near future \cite{NEAR}.

It is well known that within grand canonical ensemble the fluctuations of the condensate
are given by $\left\langle \delta ^{2}N_{\bf{0}}\right\rangle =N_{\bf{0}}\left( N_{\bf{0}}+1\right) \sim V^{2}$,
implying that $\delta N_{\bf{0}}$ becomes of order $N$ when the 
temperature approaches zero. To avoid this sort of unphysically large condensate
fluctuations, canonical (or microcanonical) ensemble has to be used to investigate 
the fluctuations of the condensate.
Within microcanonical and canonical ensemble, the fluctuations of the condensate have been
studied in a systematic way in the case of the ideal 
Bose gas \cite{BOX,ZIF,BOR,WIL,POL,GAJ,GRO,HOL2}. Recently, the question of
how interatomic interactions affect the fluctuations of the condensate has been the object
of several theoretical investigations \cite{GIO,IDZ1,IDZ2,MEI,KOC,GRA}.
Giorgini {\it et al.} \cite{GIO} 
found the anomalous behavior of the fluctuations
in a weakly interacting Bose gas confined in a box
within the traditional particle-number-nonconserving Bogoliubov approach.
In \cite{GIO} the fluctuations of the condensate follows the law 
$\left\langle \delta ^{2}N_{\bf{0}}\right\rangle \sim V^{4/3}$.
However, Idziaszek {\it et al.} \cite{IDZ1} considered that 
the fluctuations are proportional to the volume.
Recently, Kocharovsky {\it et al.} \cite{KOC} supported and extended the results of the
work of Giorgini {\it et al.} \cite{GIO} using the particle-number-conserving operator formalism.

Although the correction to the ground state occupation number due to interatomic interaction
has been clearly discussed within grand canonical ensemble \cite{GIO1}
and canonical ensemble \cite{XIONG},
the role of interaction on
the condensate fluctuations of the weakly interacting Bose gas is still an open and unsolved 
problem. Different from the ground state occupation number,
different models of describing the weakly interacting
Bose gas will lead to vastly different prediction concerning the fluctuations
of the condensate. 

The purpose of this paper is to present a unified method to
calculate the fluctuations of the condensate when different ways of approximation
to the weakly interacting Bose gases are used. Within the canonical ensemble we give the
distribution function of the ground state occupation number for the ideal and interacting 
Boson system in a box. 
We obtain the fluctuations of the condensate from the distribution function.
In particular, we found that the distribution function is not Gaussian function
in the case of interacting Boson system in a box.
The paper is organized as follows. Sec.II is devoted to outline the canonical ensemble,
which is developed to discuss the fluctuations of the condensate for the ideal Bose
gas in a box. In Sec.III we investigate the fluctuations of the interacting Bose gas
based on the lowest order perturbation theory. In Sec.IV the fluctuations are calculated
based on the Bogoliubov theory. Finally, we give a discussion and summary
of the results in Sec.V.

\section{Mean Ground State Occupation Number and Fluctuations in the Ideal Bose Gases}

Let us start our investigation on the fluctuations of the ideal Bose gases in the frame
of canonical ensemble.
According to the canonical ensemble the partition function of the $N$ non-interacting
bosons in a box is given by

\begin{equation}
{Z_{ideal}\left[ N\right] =\sum_{\Sigma N_{\bf n}=N}\exp \left[ -\beta 
\left\{ \Sigma N_{\bf n}\varepsilon _{\bf n}\right\} \right]},
\label{par1}
\end{equation}

\noindent where $N_{\bf{n}}$ and $\varepsilon _{\bf{n}}$ are the occupation numbers
and energy level of the state ${\bf{n}}=\{n_{x},n_{y},n_{z}\}$
respectively. $\beta=1/k_{B}T$. In (\ref {par1}) the energy level of the system takes the form

\begin{equation}
{\varepsilon _{\bf{n}}=\frac{\pi ^{2}\hbar ^{2}\left( n_{x}^{2}+n_{y}^{2}+n_{z}^{2}\right) }{2mL^{2}}.}
\label{level}
\end{equation}

Separating out the ground state $\bf{n}=\bf{0}$ from the state 
$\bf{n}\neq \bf{0}$, we have

\begin{equation}
{Z_{ideal}\left[ N\right] =\sum_{N_{\bf{0}}=0}^{N}
\left\{ \exp \left[ -\beta N_{\bf{0}}\varepsilon _{\bf{0}}\right] Z_{0}\left( N-N_{\bf{0}}\right) \right\},}
\label{par2}
\end{equation}

\noindent where $Z_{0}\left( N-N_{\bf{0}}\right)$
stands for the partition function of a fictitious system comprising $N-N_{\bf{0}}$ non-interacting
bosons. Assuming $A_{0}\left( N-N_{\bf{0}}\right) $ stands for the free energy of the
fictitious system,

\begin{equation}
{A_{0}\left( N-N_{\bf{0}}\right) =-k_{B}T\ln Z_{0}\left( N-N_{\bf{0}}\right)}
\label{free}
\end{equation}

\noindent From (\ref{par2}) and (\ref{free}) the partition function $Z_{ideal}\left[ N\right] $ becomes

\begin{equation}
{Z_{ideal}\left[ N\right] =\sum_{N_{\bf{0}}=0}^{N}\exp \left[ q\left( N,N_{\bf{0}}\right) \right],}
\label{par3}
\end{equation}

\noindent where $q\left( N,N_{\bf{0}}\right) =-\beta N_{\bf{0}}\varepsilon _{\bf{0}}-\beta A_{0}\left( N-N_{\bf{0}}\right) $.
Obviously $\exp \left[ q\left( N,N_{\bf{0}}\right) \right] /Z_{ideal}\left[ N\right] $
represents the probability to find $N_{\bf 0}$ atoms in the condensate. We will
give the distribution function of the ground state occupation number in the following. 

Let us first investigate the largest term in the sum of the partition function $Z_{ideal}\left[ N\right] $.
Assume the number of the condensed atoms is $N_{\bf{0}}^{p}$ in the largest term of the
partition function $Z_{ideal}\left[ N\right] $. The largest term 
$Z_{0}\left( N-N_{\bf{0}}^{p}\right) $
is determined by the 
requirement that 
$\frac{\partial }{\partial N_{\bf{0}}}q\left( N,N_{\bf{0}}\right) |_{N_{\bf{0}}=N_{\bf{0}}^{p}}=0$, {\it ie.}

\begin{equation}
{-\beta \varepsilon _{\bf{0}}-\beta \frac{\partial }{\partial N_{\bf{0}}^{p}}A_{0}\left( N-N_{\bf{0}}^{p}\right) =0}
\label{free1}
\end{equation}

The calculations of the free energy $A_{0}\left( N-N_{\bf{0}}^{p}\right) $ is nontrivial because
there is a requirement that the number of the particles is $N-N_{\bf{0}}^{p}$ in the
summation of the partition function $Z_{0}\left( N-N_{\bf{0}}^{p}\right) $.
Using the saddle-point method developed by Darwin and Fowler \cite{DAR}
it is straightforward to obtain the free
energy $A_{0}\left( N-N_{\bf{0}}^{p}\right) $ of the fictitious system.

\begin{equation}
{A_{0}\left( N-N_{\bf{0}}^{p}\right) =\left( N-N_{\bf{0}}^{p}\right) k_{B}T\ln z_{\bf{0}}^{p}-V\frac{k_{B}T}{\lambda ^{3}}g_{5/2}\left( z_{\bf{0}}^{p}\right),}
\label{free2}
\end{equation}

\noindent where $\lambda =\sqrt{2\pi \beta \hbar ^{2}/m}$ is the thermal wavelength.
$z_{\bf{0}}^{p}$ is the fugacity of the $N-N_{\bf{0}}^{p}$ non-interacting 
bosons and is determined by the equation

\begin{equation}
{N-N_{\bf{0}}^{p}=\sum_{_{\bf{n}\neq \bf{0}}}\frac{1}{\exp \left[ \varepsilon _{\bf{n}}/k_{B}T\right] \left( z_{\bf{0}}^{p}\right) ^{-1}-1}=\frac{V}{\lambda^{3} }g_{3/2}\left( z_{\bf{0}}^{p}\right).} 
\label{pro}
\end{equation}

\noindent From (\ref{free2}) and (\ref{pro}) one finds

\begin{equation}
{-\beta \frac{\partial }{\partial N_{\bf{0}}^{p}}A_{0}\left( N-N_{\bf{0}}^{p}\right) =\ln z_{\bf{0}}^{p}.}
\label{free3}
\end{equation}

\noindent Combining (\ref{free1}) and (\ref{free3}) one obtains
$\ln z_{\bf{0}}^{p}=\beta \varepsilon _{\bf{0}}$.
Therefore the most probable value $N_{\bf{0}}^{p}$ is determined by 

\begin{equation}
{N_{\bf{0}}^{p}=N-\sum_{\bf{n}\neq \bf{0}}\frac{1}{\exp \left[ \left( \varepsilon _{\bf{n}}-\varepsilon _{\bf{0}}\right) /k_{B}T\right] -1}.}
\label{roo}
\end{equation}

\noindent $N_{\bf{0}}^{p}$ is exactly the mean occupation number of the condensate atoms in
the frame of the grand canonical ensemble. For sufficiently large $N$, the sum
$\sum_{N_{\bf{0}}=0}^{N}$ in (\ref{par2}) maybe replaced by the largest term, for
the error omitted in doing so will be statistically negligible.
In this case, (\ref{roo}) shows the equivalence between canonical ensemble and 
grand canonical ensemble for large $N$. From (\ref{roo}), below the critical temperature
$N_{\bf{0}}^{p}$ is determined by

\begin{equation}
{N_{\bf{0}}^{p}=N-\frac{V}{\lambda ^{3}}\zeta \left( 3/2\right) =N\left( 1-\left( \frac{T}{T_{c}^{0}}\right) ^{3/2}\right),}
\label{most}
\end{equation}

\noindent where 
$T_{c}^{0}=\frac{2\pi }{\left[ \zeta \left( 3/2\right) \right] ^{2/3}}\frac{\hbar ^{2}}{mk_{B}}\left( \frac{N}{V}\right) ^{2/3}$
is the transition temperature of the ideal Bose gas.

Other terms in (\ref{par3}) will contribute to the fluctuations of the condensate, and
lead to the deviation of the mean occupation number $\left\langle N_{\bf{0}}\right\rangle$
from the most probable value
$N_{\bf{0}}^{p}$. When $N_{\bf{0}}\neq N_{\bf{0}}^{p}$,
$\frac{\partial }{\partial N_{\bf{0}}}q\left( N,N_{\bf{0}}\right) \neq 0$.
Assuming 

\begin{equation}
{\frac{\partial }{\partial N_{\bf{0}}}q\left( N,N_{\bf{0}}\right) =\alpha \left( N,N_{\bf{0}}\right),}
\label{qalpha}
\end{equation}

\noindent and repeating the saddle-point method, one obtains

\begin{equation}
{N_{\bf{0}}=N-\sum_{\bf{n}\neq \bf{0}}\frac{1}{\exp \left[ \left( \varepsilon _{\bf{n}}-\varepsilon _{\bf{0}}\right) /k_{B}T\right] \exp \left[ -\alpha \left( N,N_{\bf{0}}\right) \right] -1}.}
\label{non}
\end{equation}

From (\ref{most}) and (\ref{non}) one gets

\begin{equation}
{\alpha \left( N,N_{\bf{0}}\right) =-\frac{\lambda ^{6}}{V^{2}}\frac{\left( N_{\bf{0}}-N_{\bf{0}}^{p}\right) ^{2}}{4\pi }\theta \left( N_{\bf{0}}-N_{\bf{0}}^{p}\right) }
\label{alpha} 
\end{equation}

\noindent where $\theta \left( N_{\bf{0}}-N_{\bf{0}}^{p}\right) $
is a Sign function. $\theta \left( N_{\bf{0}}-N_{\bf{0}}^{p}\right) =1$ when $N_{\bf{0}}>N_{\bf{0}}^{p}$, 
and $\theta \left( N_{\bf{0}}-N_{\bf{0}}^{p}\right) =-1$ when $N_{\bf{0}}<N_{\bf{0}}^{p}$.
To obtain (\ref{alpha}) we have used the expansions $g_{3/2}\left( 1-\delta \right) \approx \zeta \left( 3/2\right) -2\sqrt{\pi \delta }$ \cite{ROB}
and the approximation $\exp \left[ -\alpha \left( N,N_{\bf{0}}\right) \right] \approx 1-\alpha \left( N,N_{\bf{0}}\right) $.
From (\ref{qalpha}) and (\ref{alpha}) one obtains easily the following result for $q\left( N,N_{\bf{0}}\right) $.

$${q\left( N,N_{\bf{0}}\right) =\int_{N_{\bf{0}}^{p}}^{N_{\bf{0}}}\alpha \left( N,N_{\bf{0}}\right) dN_{\bf{0}}
+q\left( N,N_{\bf{0}}^{p}\right)}$$

\begin{equation}
{=-\frac{\lambda ^{6}}{12\pi V^{2}}\left| N_{\bf{0}}-N_{\bf{0}}^{p}\right| ^{3}
+q\left( N,N_{\bf{0}}^{p}\right).}
\label{qqq}
\end{equation}

The partition function $Z_{ideal}\left[ N\right] $ is thus

\begin{equation}
{Z_{ideal}\left[ N\right] =\sum_{N_{\bf{0}}=0}^{N}\left\{ \exp \left[ q\left( N,N_{\bf{0}}^{p}\right) \right] G_{ideal}\left( N,N_{\bf{0}}\right) \right\},}
\label{parideal}
\end{equation}

\noindent where we have introduced  
a distribution function $G_{ideal}\left( N,N_{\bf{0}}\right) $,

\begin{equation}
{G_{ideal}\left( N,N_{\bf{0}}\right) =\exp \left[ -\frac{\lambda ^{6}}{12\pi V^{2}}\left| N_{\bf{0}}-N_{\bf{0}}^{p}\right| ^{3}\right].}
\label{Gauss}
\end{equation}

Assuming $P\left( N_{\bf{0}}|N\right) $ is the probability to find $N_{\bf{0}}$ atoms
in the condensate, the distribution function $G_{ideal}\left( N,N_{\bf{0}}\right) $
represents the ratio $\frac{P\left( N_{\bf{0}}|N\right) }{P\left( N_{\bf{0}}^{p}|N\right) }$,
{\it ie.} the relative probability to find $N_{\bf 0}$ atoms in the condensate.
From (\ref{parideal}) and (\ref{Gauss}) one obtains the mean occupation number
$\left\langle N_{\bf{0}}\right\rangle $ and fluctuations
$\left\langle \delta^{2} N_{\bf{0}}\right\rangle $ within the canonical ensemble,

\begin{equation}
{\left\langle N_{\bf{0}}\right\rangle =\frac{\sum_{N_{\bf{0}}=0}
^{N}N_{\bf{0}}G_{ideal}\left( N,N_{\bf{0}}\right) }{\sum_{N_{\bf{0}}=0}^{N}G_{ideal}\left( N,N_{\bf{0}}\right) },}
\label{meanideal}
\end{equation}

$${\left\langle \delta^{2} N_{\bf{0}}\right\rangle =\left\langle N_{\bf{0}}^{2}\right\rangle 
-\left\langle N_{\bf{0}}\right\rangle ^{2}=}$$

\begin{equation}
{\frac{\sum_{N_{\bf{0}}=0}^{N}N_{\bf{0}}^{2}
G_{ideal}\left( N,N_{\bf{0}}\right) }{\sum_{N_{\bf{0}}=0}^{N}
G_{ideal}\left( N,N_{\bf{0}}\right) }-\left( \frac{\sum_{N_{\bf{0}}=0}^{N}N_{\bf{0}}
G_{ideal}\left( N,N_{\bf{0}}\right) }{\sum_{N_{\bf{0}}=0}^{N}
G_{ideal}\left( N,N_{\bf{0}}\right) }\right) ^{2}.}
\label{fluc}
\end{equation}

\noindent From (\ref{most}), (\ref{Gauss}) and (\ref{meanideal}), (\ref{fluc}) it is easy to obtain
$\left\langle N_{\bf{0}}\right\rangle$
and
$\left\langle \delta^{2} N_{\bf{0}}\right\rangle $ of the non-interacting 
Bose gases in a box. At the critical temperature $T_{c}^{0}$, $N_{\bf{0}}^{p}=0$.
Thus $G_{ideal}\left( T=T_{c}^{0}\right) =\exp \left[ -\frac{\lambda_{0} ^{6}}{12\pi V^{2}}N_{\bf{0}}^{3}\right] $, where
$\lambda _{0}$ is the thermal wavelength at $T_{c}^{0}$.
From (\ref{meanideal}) and (\ref{fluc}) one obtains the analytical result for the condensate
fluctuations at $T_{c}^{0}$.

$${\left\langle \delta N_{\bf{0}}^{2}\right\rangle _{T=T_{c}^{0}}=}$$

\begin{equation}
{\left[ \frac{1}{3\Gamma \left( 4/3\right) }-\left( \frac{\Gamma \left( 5/3\right) }{2\Gamma \left( 4/3\right) }\right) ^{2}\right] \left( \frac{12\pi }{\lambda _{0}^{6}}\right) ^{2/3}V^{4/3},}
\label{nearideal}
\end{equation}

\noindent where $\Gamma \left( n\right) =\int_{0}^{\infty }e^{-t}t^{n-1}dt$ is Gamma function.
$\Gamma \left( 4/3\right) =0.893$ and $\Gamma \left( 5/3\right) =0.903$.
(\ref{nearideal}) clearly shows that there is anomalous behavior for 
the fluctuations of the condensate. 
When $T\rightarrow 0$, from (\ref{Gauss}) one
finds $G_{ideal}\left( N,N_{\bf{0}}\right) \rightarrow 0$ when $N_{\bf{0}}\neq N$. Therefore
when  $T\rightarrow 0$ one obtains $\left\langle N_{\bf{0}}\right\rangle \rightarrow N$
and $<\delta ^{2}N_{\bf{0}}>\rightarrow 0$. 

In Fig.1 we plot $\left\langle N_{\bf{0}}\right\rangle /N$ 
as a function of temperature
for the ideal Bose gases in a box. The solid line displays the mean ground state occupation number
within the grand canonical ensemble (or $N_{\bf{0}}^{p}$).
When $N>10^{4}$, the mean ground state occupation number of the canonical ensemble
agrees well with that of the grand canonical ensemble.
Obviously, in the case of $N\rightarrow \infty $, the mean ground state occupation number of
the canonical ensemble coincides with that of the grand canonical ensemble.

\begin{figure}[tb]
\psfig{figure=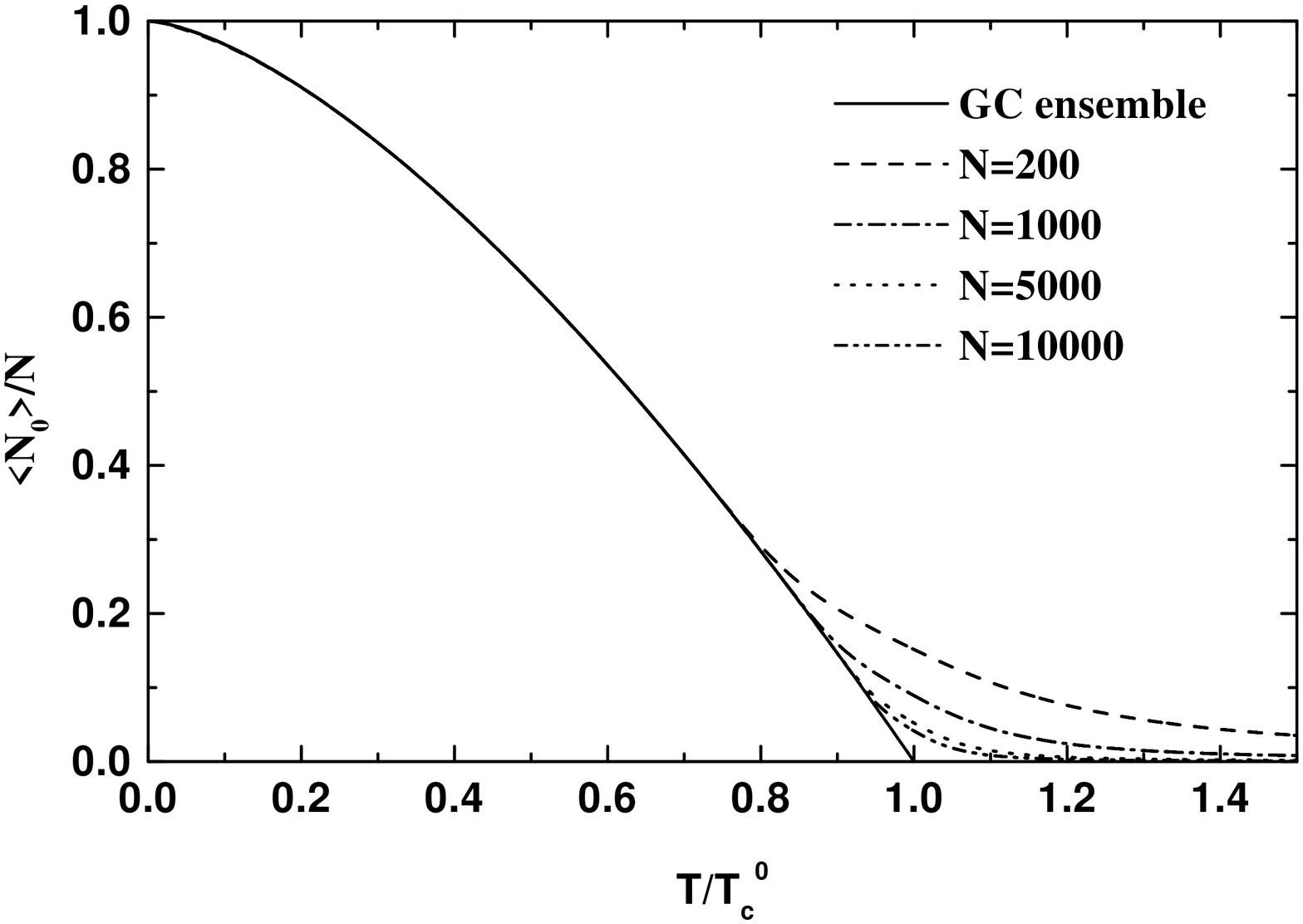,width=\columnwidth,angle=270}
\caption{Temperature dependence of the mean ground state occupation number
for ideal Bose gas confined in a box within the canonical ensemble. The solid line
shows $\left\langle N_{\bf{0}}\right\rangle /N$ within the grand canonical ensemble (or
$N_{\bf{0}}^{p}/N$ in the canonical ensemble). When
$N\rightarrow \infty $, the mean ground state occupation number of the canonical ensemble
coincides with that of the grand canonical ensemble.} 
\end{figure}

\begin{figure}[tb]
\psfig{figure=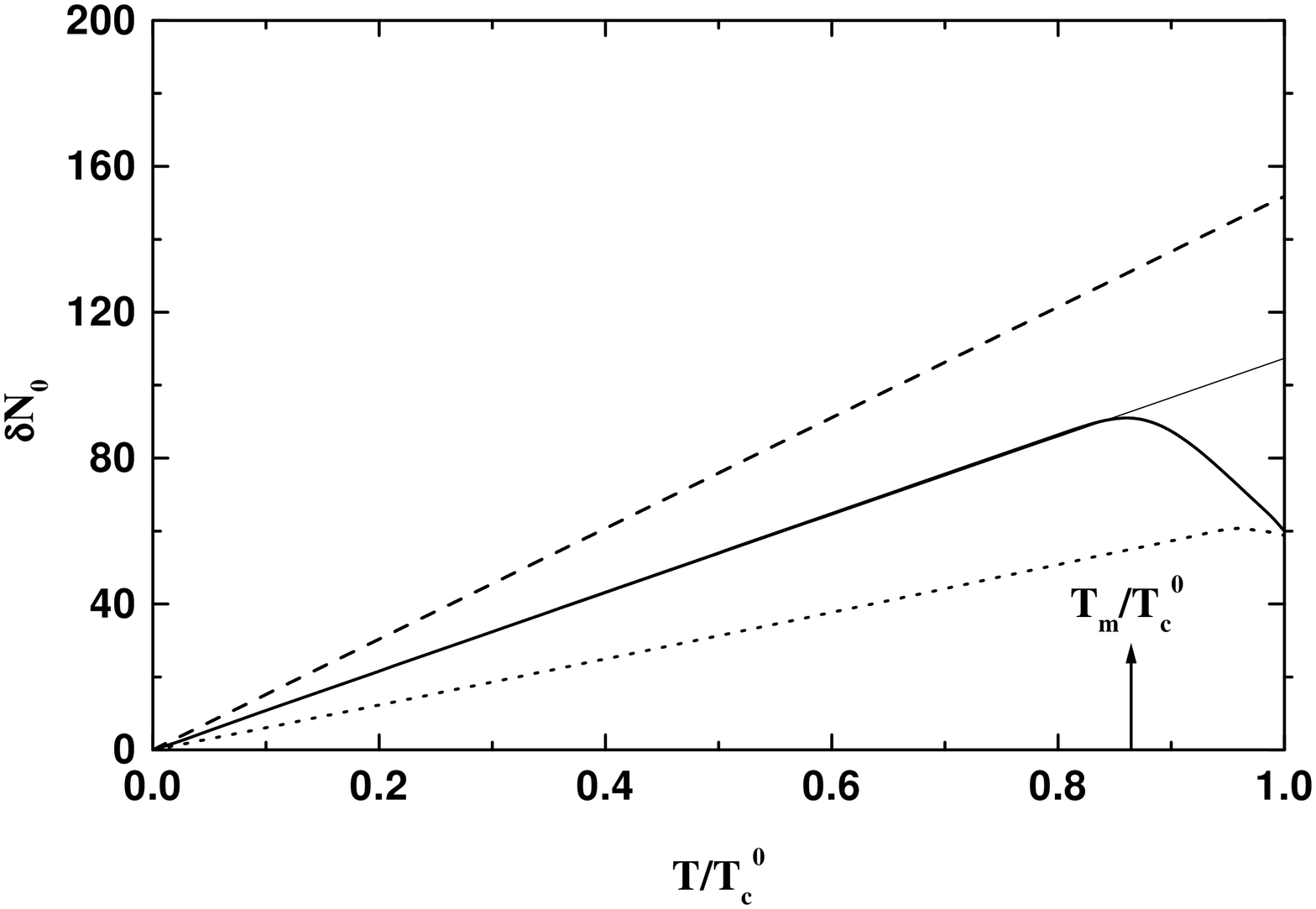,width=\columnwidth,angle=270}
\caption{Temperature dependence of $\delta N_{0}$ for the ideal Bose gas confined in a
box. The thick solid line displays the numerical result of (\ref{fluc}), while the thin solid
line shows the analytical result (\ref{belowbelow}) for $\delta N_{0}$ below $T_{m}$ (The arrow
marks $T_{m}$ which corresponds to the maximum condensate fluctuations).
The dashed line is obtained from (\ref{idealgio}) which comes from [14]. The dotted line
displays the numerical result of Wilkens {\it et al.} [9].}
\end{figure}

In Fig.2 we plot numerical result of
$\delta N_{\bf{0}}=\sqrt{\left\langle \delta^{2} N_{\bf{0}}\right\rangle }$ (thick solid line) 
for the ideal Bose gas with  
$N=1000$ atoms confined in a three-dimensional box. 
The dashed line shows the result of \cite{GIO}, where the fluctuations is given by

\begin{equation}
{\left\langle \delta^{2} N_{\bf{0}}\right\rangle =
2A\left( \frac{mk_{B}T}{\hbar ^{2}}\right) ^{2}V^{4/3}.}
\label{idealgio}
\end{equation}

\noindent The coefficient in (\ref{idealgio}) is 
$A=2/\pi ^{4}\times \Sigma _{{\bf{n}}\neq {\bf{0}}}1/{\bf{n}}^{4}=0.105$. The dashed line is larger than our
result because of the approximation in \cite{GIO}. In \cite{GIO}
$\left\langle \delta^{2} N_{\bf{0}}\right\rangle =\sum_{\bf{n}\neq \bf{0}}f_{\bf{n}}^{2}$,
where $f_{\bf{n}}=\left( \exp \left( \varepsilon _{\bf{n}}/k_{B}T\right) -1\right) ^{-1}$.
For the convenience of calculations, $f_{\bf{n}}$ is approximated as 
$\varepsilon _{\bf{n}}/k_{B}T$ for low energy atoms. However, this approximation is used 
also for the atoms whose energy level is larger than $k_{B}T$. Obviously, this approximation
will lead to the fluctuations become larger.
On the other hand, (\ref{idealgio}) holds in the canonical ensemble except near and
above $T_{c}$, while our analysis holds also for the temperature near $T_{c}^{0}$. In Fig.2 
the dotted line shows
the numerical result of Wilkens {\it et al.} \cite{WIL}.

In Fig.2 the arrow marks the temperature $T_{m}$ which corresponds to the maximum
fluctuations $\left\langle \delta ^{2}N_{\bf{0}}\right\rangle _{\max }$. Below the temperature
$T_{m}$, from (\ref{fluc}), one obtains the analytical result for the fluctuations of the
condensate.

\begin{equation}
{\left\langle \delta ^{2}N_{\bf{0}}\right\rangle =A\left( \frac{m{k_{B}T}}{\hbar ^{2}}\right) ^{2}V^{4/3}.}
\label{belowbelow}
\end{equation}

\noindent It is interested to find that the coefficient differs by a factor $2$, compared
to (\ref{idealgio}). The thin solid line shows (\ref{belowbelow}) in Fig.2.

We should note that our results are reliable although the disputable saddle-point method
is used to investigate the fluctuations of the condensate. It is well known that
the applicability of the saddle-point approximation for the condensed Bose gases
has been the subject of a long debate \cite{ZIF,DEB}. Recently, the analysis in \cite{GRO}
showed that the fluctuations are overestimated, and do not appear to vanish properly with
temperature using the usual saddle-point method. Our discussions for fluctuations are reasonable
because of two reasons. (i) As proved in \cite{HOL2}, the free energy (\ref{free2}) of the
non-interacting Bose gases is still correct, even when carefully
deal with the failure of the standard saddle-point method below the critical temperature.
(ii) In the usual statistical method
$\left\langle N_{\bf{0}}\right\rangle $ and $\left\langle \delta ^{2}N_{\bf{0}}\right\rangle $
are obtained through the first and second partial derivative of the partition function respectively.
When  saddle-point approximation is used to calculate the partition function of the system,
the error will be overestimated in the second partial derivative of the partition function
so that we can not obtain correct fluctuations of the condensate in the usual method.
However, in this paper what we used is the reliable result (\ref{free2}).
The distribution function of the ground state occupation number is obtained
directly from (\ref{free2}), without resorting to the second partial derivative
of the partition function. $\left\langle N_{\bf{0}}\right\rangle $ and 
$\left\langle \delta ^{2}N_{\bf{0}}\right\rangle $ are obtained from the distribution function
in this paper.

\section{Fluctuations of the Condensate Based on the Lowest Order Perturbation Theory}

In the case of interacting Bose gases, the role of interactions on the fluctuations of the
condensate is still an open and unsolved problem. Giorgini {\it et al.} \cite{GIO} predicted
the anomalous behavior of the fluctuations in a weakly interacting Bose gas confined
in a box, while Idziaszek {\it et al.} \cite{IDZ1} considered that the fluctuations are normal. Researches
show that different model of approximation to the interacting Bose gases will
lead to different predictions concerning the fluctuations of the condensate.
The method developed to obtain the fluctuations of the ideal Bose gas in this paper
can be used straightforwardly to discuss the fluctuations of the
interacting Bose gases when different models of approximation are adopted.

Let us first discuss the fluctuations of the condensate
in the case of the lowest order
perturbation theory, which is also discussed in \cite{IDZ1}. In terms of the lowest order perturbation
theory, the partition function of the system within the canonical ensemble is given by

\begin{equation}
{Z_{int}\left[ N\right] =\sum_{\Sigma N_{\bf{n}}=N}\exp \left[ -\beta \left( \Sigma N_{\bf{n}}\varepsilon _{\bf{n}}+E_{int}\right) \right],}
\label{intpar}
\end{equation}

\noindent where the interaction energy of the system takes the form \cite{PAR,HUANG}

\begin{equation}
{E_{int}=\frac{4\pi a\hbar ^{2}}{mV}\left( N^{2}-\frac{1}{2}N_{\bf{0}}^{2}\right).}
\label{intenergy}
\end{equation}

\noindent In (\ref{intenergy}) $a$ is the scattering length. Separating out the ground state
$\bf{n}=\bf{0}$ from the state $\bf{n}\neq \bf{0}$, 
one obtains the following form for the partition function

\begin{equation}
{Z_{int}\left[ N\right] =\sum_{N_{\bf{0}}=0}^{N}\left\{ \exp \left[ -\beta N_{\bf 0}\varepsilon_{\bf 0}
-\beta E_{int}\right] Z_{0}\left( N-N_{\bf{0}}\right) \right\},}
\label{intpar1}
\end{equation}

\noindent where $Z_{0}\left( N-N_{\bf{0}}\right) $ stands for the partition function of a fictitious
$N-N_{\bf{0}}$ non-interacting bosons. 
Using the free energy $A_{0}\left( N-N_{\bf{0}}\right) $
of the fictitious system, the partition function is thus

\begin{equation}
{Z_{int}\left[ N\right] =\sum_{N_{\bf{0}}=0}^{N}\exp \left[ q\left( N,N_{\bf{0}}\right) \right],}
\label{intpar2}
\end{equation}

\noindent where $q\left( N,N_{\bf{0}}\right) $ takes the form

\begin{equation}
{q\left( N,N_{\bf{0}}\right) =-\beta N_{\bf 0}\varepsilon_{\bf 0}-\beta E_{int}-\beta A_{0}\left( N-N_{\bf{0}}\right).}
\label{intq}
\end{equation}

Analogously to the case of ideal Bose gases, let us first investigate the 
largest term in the sum of $Z_{int}\left[ N\right] $. The largest term is determined by the
requirement $\frac{\partial }{\partial N_{\bf{0}}}q\left( N,N_{\bf{0}}\right) |_{N_{\bf{0}}=N_{\bf{0}}^{p}}=0$.
Therefore one obtains the most probable value
$N_{\bf{0}}^{p}$ of the interacting Bosons.

\begin{equation}
{N_{\bf{0}}^{p}=N-\sum_{\bf{n}\neq \bf{0}}\frac{1}{\exp \left[ \beta \varepsilon _{\bf{n}}\right] \left( z_{\bf{0}}^{p}\right) ^{-1}-1},}
\label{intermost}
\end{equation}

\noindent where $z_{\bf{0}}^{p}$ is determined by

\begin{equation}
{ln z_{\bf{0}}^{p}=\beta \varepsilon_{\bf 0}+\beta \frac{\partial }{\partial N_{\bf{0}}^{p}}E_{int}=
\beta \varepsilon_{\bf 0}-\frac{2a\lambda ^{2}N_{\bf{0}}^{p}}{V}}
\label{zzz1}
\end{equation}

\noindent From (\ref{intermost}) and (\ref{zzz1}) one obtains

\begin{equation}
{N_{\bf{0}}^{p}\simeq N-\frac{V}{\lambda ^{3}}\left[ \zeta \left( 3/2\right) -2\sqrt{\pi }\left( \frac{2a\lambda ^{2}N_{\bf{0}}^{p}}{V}\right) ^{1/2}\right] }
\label{mostinter}
\end{equation}

Other terms in (\ref{intpar2}) will contribute to the fluctuations of the system.
Assuming $\frac{\partial }{\partial N_{\bf{0}}}q\left( N,N_{\bf{0}}\right) =\alpha \left( N,N_{\bf{0}}\right) $,
one obtains the result for $N_{\bf{0}}$

\begin{equation}
{N_{\bf{0}}=N-\sum_{\bf{n}\neq \bf{0}}\frac{1}{\exp \left[ \beta \varepsilon _{\bf{n}}\right] \left( z_{\bf{0}}\right) ^{-1}-1},}
\label{other}
\end{equation}

\noindent where $z_{\bf{0}}$ is determined by

\begin{equation}
{\ln z_{\bf{0}}=\beta \varepsilon_{\bf 0}-\frac{2a\lambda ^{2}N_{\bf{0}}}{V}+\alpha \left( N,N_{\bf{0}}\right)}
\label{zzzz}
\end{equation}

From (\ref{intermost}), (\ref{zzz1}) and (\ref{other}), (\ref{zzzz}) it is straightforward to obtain
the distribution function $G_{int}\left( N,N_{\bf{0}}\right) $
of the interacting Bose gases.

\begin{equation}
{G_{int}\left( N,N_{\bf{0}}\right) =G_{ideal}\left( N,N_{\bf{0}}\right)
 \times R_{int}\left( N,N_{\bf{0}}\right),}
\label{intergauss}
\end{equation}

\noindent where $G_{ideal}\left( N,N_{\bf{0}}\right) $ is the distribution function
(\ref{Gauss}) of the ideal Bose gases, while $R_{int}\left( N,N_{\bf{0}}\right) $
takes the form

\begin{equation}
{R_{int}\left( N,N_{\bf{0}}\right) =
R_{1}\left( N,N_{\bf{0}}\right) \times 
R_{2}\left( N,N_{\bf{0}}\right) \times 
R_{3}\left( N,N_{\bf{0}}\right).}
\label{Rfunction}
\end{equation}

\noindent In (\ref{Rfunction}),

$${R_{1}\left( N,N_{\bf{0}}\right) =}$$

\begin{equation}
{\exp \left[ -\frac{\left( \zeta \left( 3/2\right) \right) ^{3/2}}
{\sqrt{2\pi }}\left( \frac{a}{\lambda _{0}}\frac{N_{\bf{0}}^{p}}{N}\right) ^{1/2}
\frac{\left( N_{\bf{0}}-N_{\bf{0}}^{p}\right) ^{2}}{Nt^{2}}\theta \left( N_{\bf{0}}-
N_{\bf{0}}^{p}\right) \right],}
\label{R1}
\end{equation}

\begin{equation}
{R_{2}\left( N,N_{\bf{0}}\right) =
\exp \left[ \frac{\zeta \left( 3/2\right) a}{\lambda _{0}}\frac{N_{\bf{0}}^{2}-
\left( N_{\bf{0}}^{p}\right) ^{2}}{Nt}\right],}
\label{R2}
\end{equation}

\begin{equation}
{R_{3}\left( N,N_{\bf{0}}\right) =
\exp \left[ -\left| \frac{2\zeta \left( 3/2\right) a}{\lambda _{0}}
\frac{N_{\bf{0}}^{p}\left( N_{\bf{0}}-N_{\bf{0}}^{p}\right) }{Nt}\right| \right].}
\label{R3}
\end{equation}

We should note that 
$G_{int}\left( N,N_{\bf{0}}\right) $ is not a Gaussian distribution function because of
the non-Gaussian factors $R_{1}\left( N,N_{\bf{0}}\right) $ and $R_{2}\left( N,N_{\bf{0}}\right) $,
while Idziaszek {\it et al.} \cite{IDZ1} utilized the 
Gaussian distribution as an assumption to investigate the fluctuations of the interacting 
system. In $R_{int}\left( N,N_{\bf{0}}\right) $,
$R_{1}\left( N,N_{\bf{0}}\right) $ comprises the factor $\left( a/\lambda _{0}\right) ^{1/2}$
and represents the leading correction to the distribution function due to interatomic interaction,
while $R_{2}\left( N,N_{\bf{0}}\right) $ and $R_{3}\left( N,N_{\bf{0}}\right) $
are high order correction to the distribution function.
We should note that the leading contribution $R_{1}\left( N,N_{\bf{0}}\right) $
is not a Gaussian function.

From the distribution function 
$G_{int}\left( N,N_{\bf{0}}\right) $
the mean occupation number and 
fluctuations of the condensate are determined by

\begin{equation}
{\left\langle N_{\bf{0}}\right\rangle =\frac{\sum_{N_{\bf{0}}=0}^{N}N_{\bf{0}}G_{int}\left( N,N_{\bf{0}}\right) }{\sum_{N_{\bf{0}}=0}^{N}G_{int}\left( N,N_{\bf{0}}\right) },}
\label{intermean}
\end{equation}

$${\left\langle \delta^{2} N_{\bf{0}}\right\rangle =}$$

\begin{equation}
{\frac{\sum_{N_{\bf{0}}=0}^{N}N_{\bf{0}}^{2}G_{int}\left( N,N_{\bf{0}}\right) }{\sum_{N_{\bf{0}}=0}^{N}G_{int}\left( N,N_{\bf{0}}\right) }-\left( \frac{\sum_{N_{\bf{0}}=0}^{N}N_{\bf{0}}G_{int}\left( N,N_{\bf{0}}\right) }{\sum_{N_{\bf{0}}=0}^{N}G_{int}\left( N,N_{\bf{0}}\right) }\right) ^{2}.}
\label{interflu}
\end{equation}

\begin{figure}[tb]
\psfig{figure=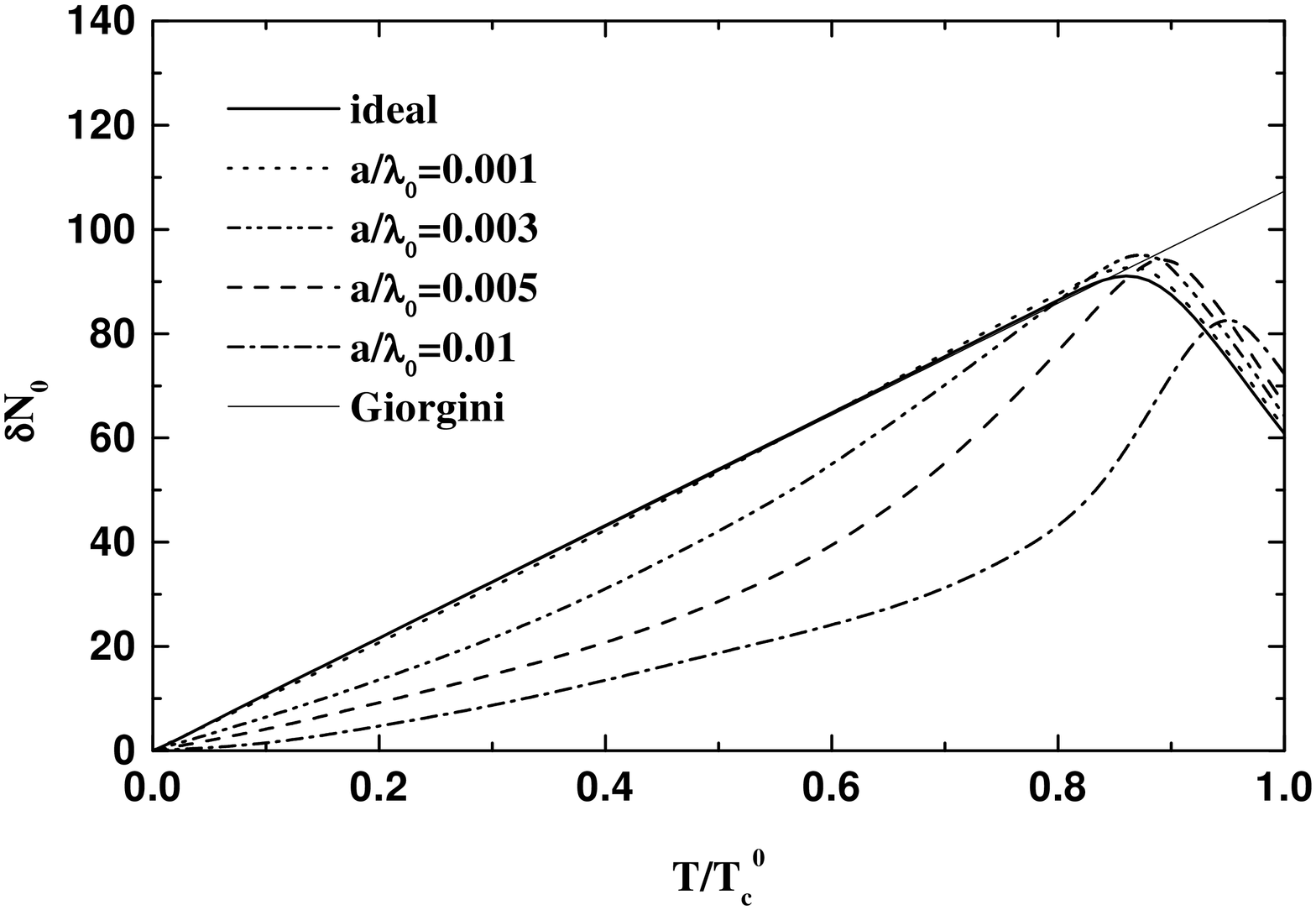,width=\columnwidth,angle=270}
\caption{Temperature dependence of $\delta N_{0}$ for interacting Bose gases based on
the lowest order perturbation theory. The thick solid line displays the numerical result of
the ideal Bose gas. We give the numerical result for the repulsive interactions with
$a/\lambda _{0}=1\times 10^{-3}$, $2\times 10^{-3}$, $5\times 10^{-3}$, $1\times 10^{-2}$.
The crossover from interacting to non-interacting Bose gases is clearly demonstrated.
The thin solid line is obtained from (8) in [14].}
\end{figure}

From (\ref{mostinter}), (\ref{intergauss}) and (\ref{intermean}), (\ref{interflu})
we can obtain the fluctuations of the interacting Boson gases.
In Fig.3 we give the numerical result for the repulsive interactions with
$a/\lambda _{0}=1\times 10^{-3}$, $2\times 10^{-3}$,
$5\times 10^{-3}$, $1\times 10^{-2}$. The crossover from interacting to ideal Bose gases
(thick solid line) is clearly demonstrated in Fig.3, while in \cite{GIO}
the fluctuations (thin solid line) of the interacting Bose gases are irrelative
to the scattering length. 
When $a\rightarrow 0$ it is easy to recover the fluctuations of the ideal Bose gases.
According to (\ref{interflu}) the leading contributions to
the fluctuations are anomalous, {\it ie.} proportional to $N^{4/3}$, while
there are also normal contributions proportional to $N$ due to interatomic interaction.
This conclusion
contradicts with that of \cite{IDZ1} which predicts normal behavior, where the lowest order
perturbation theory is also used to discuss the fluctuations of the condensate.

\section{Fluctuations of the Condensate Based on Bogoliubov Theory}

Let us investigate the fluctuations of the condensate in the framework of Bogoliubov
theory of a uniform weakly interacting Bose gas confined in a box. According to 
Bogoliubov theory \cite{BOG,GIO2}, the total number of particles out of the condensate is given by

\begin{equation}
{N_{T}=\sum_{\bf{n}\neq \bf{0}}N_{\bf{n}}=\sum_{\bf{n}\neq \bf{0}}\left( u_{\bf{n}}^{2}+v_{\bf{n}}^{2}\right) f_{\bf{n}},}
\label{bthermal}
\end{equation}

\noindent where

\begin{equation}
{u_{\bf{n}}^{2}+v_{\bf{n}}^{2}=\frac{\left( \left( \varepsilon _{\bf{n}}^{B}\right) ^{2}+g^{2}n_{0}^{2}\right) ^{1/2}}{2\varepsilon _{\bf{n}}^{B}},}
\label{ufunction}
\end{equation}

\begin{equation}
{u_{\bf{n}}v_{\bf{n}}=-\frac{gn_{0}}{2\varepsilon _{\bf{n}}^{B}},}
\label{vfunction}
\end{equation}

\noindent and $f_{\bf{n}}$ is the number of quasi-particles present in the system
at the thermal equilibrium.

\begin{equation}
{f_{\bf{n}}=\frac{1}{\exp \left[ \varepsilon _{\bf{n}}^{B}/k_{B}T\right] -1}}
\label{bdistribution}
\end{equation}

\noindent In addition, the energy of the quasi-particles entering (\ref{ufunction}) and (\ref{vfunction})
is given by the well known Bogoliubov spectrum

\begin{equation}
{\varepsilon _{\bf{n}}^{B}=\left( \left( \varepsilon _{\bf{n}}+gn_{0}\right) ^{2}-g^{2}n_{0}^{2}\right) ^{1/2},}
\label{benergy}
\end{equation}

\noindent where $g=4\pi \hbar ^{2}a/m$ is the coupling constant, and $n_{0}=N_{0}/V$
is the condensate density. At low $\left| \bf{n}\right| =\sqrt{n_{x}^{2}+n_{y}^{2}+n_{z}^{2}}$,
one obtains $u_{\bf{n}}^{2}\simeq v_{\bf{n}}^{2}\propto 1/\left| \bf{n}\right| $
and $f_{\bf{n}}\propto 1/\left| \bf{n}\right| $. This results in $1/\left| \bf{n}\right| ^{2}$
divergence in (\ref{bdistribution}) at low $\left| \bf{n}\right| $. Although this sort of divergence will not lead to large
contribution to the number of low energy quasi-particles, it gives leading
contribution to the fluctuations of the condensate, as pointed out in \cite{GIO}.
We will investigate the fluctuations due to low energy quasi-particles in the following.

In (\ref{bthermal}) $N_{\bf{n}}$ can be regarded as the effective occupation number
of the thermal atoms, while 

\begin{equation}
{N_{\bf{n}}^{B}=\frac{N_{\bf{n}}}{u_{\bf{n}}^{2}+v_{\bf{n}}^{2}}=f_{\bf{n}}}
\label{bogo}
\end{equation}

\noindent is the occupation number of the quasi-particles. From the form of 
$f_{\bf{n}}$, we can construct the partition function of the quasi-particles in the
frame of canonical ensemble.

\begin{equation}
{Z_{B}=\sum_{\left\{ \bf{n}\right\} }\exp \left[ -\beta \sum_{\bf{n}}N_{\bf{n}}^{B}\varepsilon _{\bf{n}}^{B}\right].}
\label{bpar1}
\end{equation} 

\noindent From (\ref{bogo}) $Z_{B}$ becomes

\begin{equation}
{Z_{B}=\sum_{\left\{ \Sigma N_{\bf{n}}=N\right\} }\exp \left[ -\beta \sum_{\bf{n}}N_{\bf{n}}\varepsilon _{\bf{n}}^{eff}\right],}
\label{bpar2}
\end{equation}

\noindent where $\varepsilon _{\bf{n}}^{eff}=\varepsilon _{\bf{n}}^{B}/\left( u_{\bf{n}}^{2}+v_{\bf{n}}^{2}\right) $
can be regarded as an effective energy level of the thermal atoms. In this case 
$Z_{B}$ is the partition function of a fictitious Boson system comprising $N$
non-interacting Bosons whose energy level is determined by $\varepsilon _{\bf{n}}^{eff}$.
From (\ref{bpar2}) the most probable value $N_{\bf{0}}^{p}$ is given by

\begin{equation}
{N_{\bf{0}}^{p}=N-\sum_{\bf{n}\neq \bf{0}}\frac{1}{\exp \left[ \left( \varepsilon _{\bf{n}}^{eff}-\varepsilon _{\bf{0}}^{eff}\right) /k_{B}T\right] -1},}
\label{bmost}
\end{equation}

\noindent Obviously the occupation number of low $\left| \bf{n}\right| $ in (\ref{bmost})
coincides with that of (\ref{bthermal}). Analogously, other $N_{\bf{0}}$ is thus

$${N_{\bf{0}}=N-}$$

\begin{equation}
{\sum_{\bf{n}\neq \bf{0}}\frac{1}{\exp \left[ \left( \varepsilon _{\bf{n}}^{eff}-\varepsilon _{\bf{0}}^{eff}\right) /k_{B}T\right] \exp \left[ -\alpha \left( N,N_{\bf{0}}\right) \right] -1}.}
\label{bother}
\end{equation}

From (\ref{bmost}) and (\ref{bother}) one gets

\begin{equation}
{\alpha \left( N,N_{\bf{0}}\right) \approx -\frac{N_{\bf{0}}-N_{\bf{0}}^{p}}{\sum_{\bf{n}\neq \bf{0}}\left( u_{\bf{n}}^{2}+v_{\bf{n}}^{2}\right) ^{2}f_{\bf{n}}^{2}},}
\label{balpha}
\end{equation}

\noindent where we have used the approximation 
$f_{\bf{n}}\approx k_{B}T/\varepsilon _{\bf{n}}^{B}$
for low energy quasi-particles. Therefore the Gaussian distribution function of the system
is given by

$${G_{B}\left( N,N_{\bf{0}}\right) 
=\exp \left[ -\frac{\left( N_{\bf{0}}-N_{\bf{0}}^{p}\right) ^{2}}
{2\sum_{\bf{n}\neq \bf{0}}\left( u_{\bf{n}}^{2}+
v_{\bf{n}}^{2}\right) ^{2}f_{\bf{n}}^{2}}\right]}$$

\begin{equation}
{ \approx \exp
 \left[ -\frac{\left( \zeta \left( 3/2\right) \right) ^{4/3}
\left( N_{\bf{0}}-N_{\bf{0}}^{p}\right) ^{2}}{\left( 2\pi \right) ^{2}AN^{4/3}t^{2}}\right] .}
\label{bgauss}
\end{equation}

\noindent Obviously the mean occupation number $\left\langle N_{\bf{0}}\right\rangle $
and fluctuations $\left\langle \delta ^{2}N_{\bf{0}}\right\rangle $ is given by

\begin{equation}
{\left\langle N_{\bf{0}}\right\rangle =\frac{\sum_{N_{\bf{0}}=0}^{N}N_{\bf{0}}G_{B}\left( N,N_{\bf{0}}\right) }{\sum_{N_{\bf{0}}=0}^{N}G_{B}\left( N,N_{\bf{0}}\right) },}
\label{bgauss1}
\end{equation}

$${\left\langle \delta^{2} N_{\bf{0}}\right\rangle =}$$

\begin{equation}
{\frac{\sum_{N_{\bf{0}}=0}^{N}N_{\bf{0}}^{2}G_{B}\left( N,N_{\bf{0}}\right) }{\sum_{N_{\bf{0}}=0}^{N}G_{B}\left( N,N_{\bf{0}}\right) }-\left( \frac{\sum_{N_{\bf{0}}=0}^{N}N_{\bf{0}}G_{B}\left( N,N_{\bf{0}}\right) }{\sum_{N_{\bf{0}}=0}^{N}G_{B}\left( N,N_{\bf{0}}\right) }\right) ^{2}.}
\label{bgauss2}
\end{equation}

From (\ref{bgauss}) and (\ref{bgauss1}), (\ref{bgauss2}) one obtains the fluctuations
of the condensate based on Bogoliubov theory. At the critical temperature, 
$G_{B}\left( T=T_{c}\right) =\exp \left[ -N_{\bf{0}}^{2}/\theta \right] $, where
$\theta =2\sum_{\bf{n}\neq \bf{0}}\left( u_{\bf{n}}^{2}+v_{\bf{n}}^{2}\right) ^{2}f_{\bf{n}}^{2}=\left( 2\pi \right) ^{2}AN^{4/3}/\left( \zeta \left( 3/2\right) \right) ^{4/3}$.
In this case, we obtains the analytical result of the condensate fluctuations.

\begin{equation}
{\left\langle \delta ^{2}N_{\bf{0}}\right\rangle _{T=T_{c}}=\left( \frac{1}{2}-\frac{1}{\pi }\right) \theta =\left( \frac{1}{2}-\frac{1}{\pi }\right) \frac{\left( 2\pi \right) ^{2}A}{\left( \zeta \left( 3/2\right) \right) ^{4/3}}N^{4/3}.}
\label{bnear}
\end{equation}

\noindent (\ref{bnear}) clearly shows that the anomalous behavior of the condensate 
fluctuations originates
from the low energy quasi-particles, which gives the anomalous factor $N^{4/3}$
through $\theta$. In Fig.4 the solid line displays our results based on the Bogoliubov theory, while
the dashed line shows the result of \cite{GIO}.

\begin{figure}[tb]
\psfig{figure=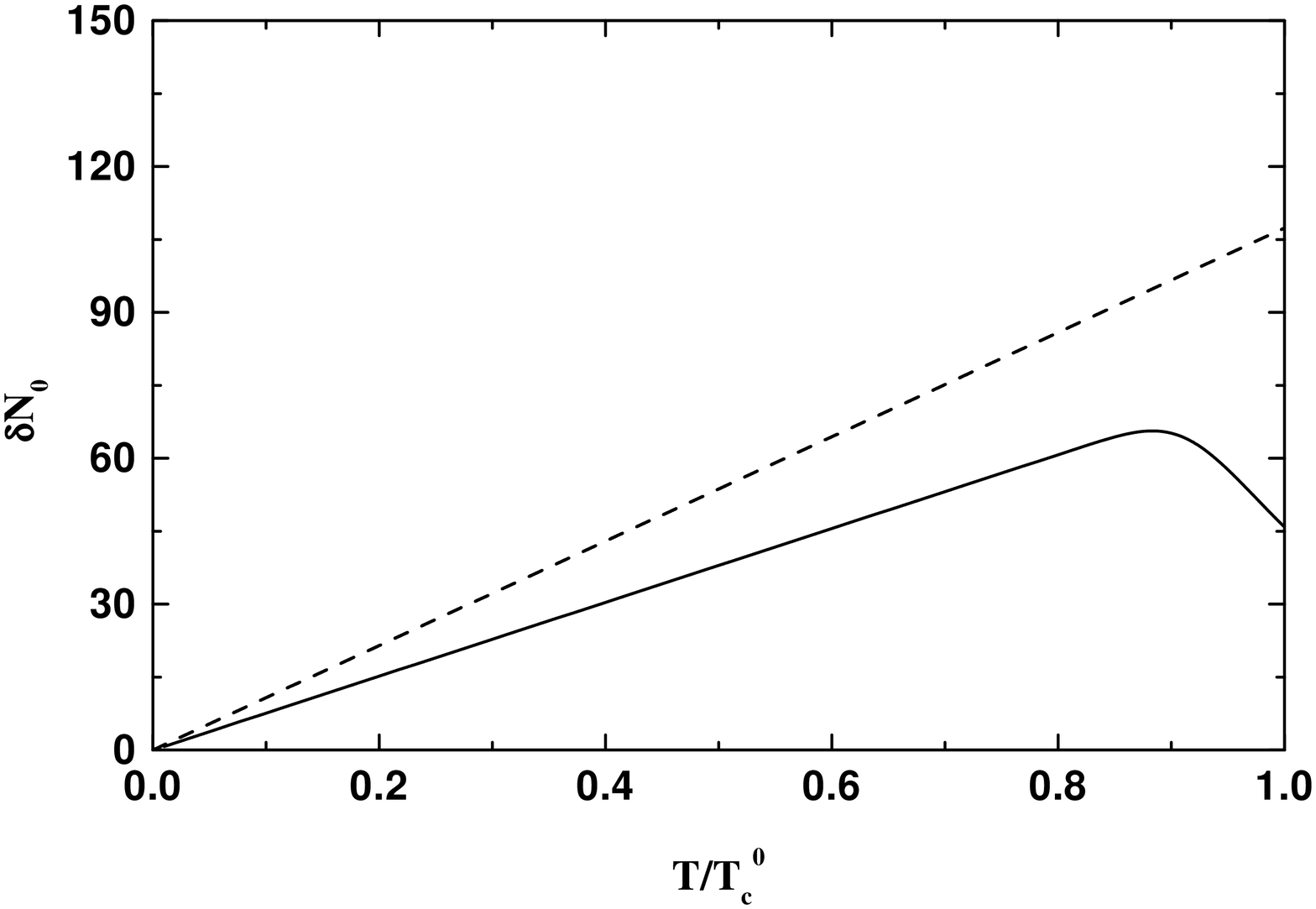,width=\columnwidth,angle=270}
\caption{Temperature dependence of $\delta N_{0}$ for interacting Bose gases based on
Bogoliubov theory. The solid line is obtained from the numerical result of (\ref{bgauss2}),
while the dashed line displays (8) in [14].}
\label{omega_ani}
\end{figure}

\section{Discussion and Conclusion}

In this paper we investigate the fluctuations of the condensate in a weakly interacting
Bose gas confined in a box. Canonical ensemble is developed to calculate the fluctuations of
the condensate when different models of interacting Bose gases are used.
We found that both the lowest order perturbation theory and Bogoliubov theory
give anomalous behavior of the fluctuations for the interacting Bose gases confined in 
a box.

Different from the usual method, the distribution function 
$P\left( N_{\bf{0}}|N\right) /P\left( N_{\bf{0}}^{p}|N\right)$ ({\it ie.} the ratio of
the probability between $N_{\bf{0}}$ and the most probable value $N_{\bf{0}}^{p}$ )
of the ground state occupation number is obtained directly to calculate 
the fluctuations of the condensate. From some senses, we give a simple method
to recover the applicability of the saddle-point approximation to discuss the condensate
fluctuations, through the avoidance of the second derivative in the usual method.

 For the present experiments of BEC, the harmonically trapped atoms are in a
situation of almost complete isolation with the outer environment surrounding 
the trap, therefore canonical (or microcanonical) ensemble should be used
to calculate the fluctuations of the condensate. On the other hand, one obtains
more accurate mean ground state occupation number within the canonical ensemble,
compared to the grand canonical ensemble. The present work may serve as another
method to investigate the thermodynamic properties  of the harmonically trapped interacting
Bose gases such as critical temperature, condensate fraction, and fluctuations of the
condensate.

The remain challenge is to extend the idea of this paper to the case of microcanonical
ensemble where the energy of the system is also invariant. In addition, the role of 
interactions on the fluctuations of the condensate are expected to be much more
dramatic in the case of attractive forces. We will investigate these problems in
a subsequent work.

\section*{Acknowledgments}
This work was supported by the Science Foundation of
Zhijiang College, Zhejiang University of Technology. This work was also
supported by the National Natural Science Foundation of China.


\begin{references}

\bibitem{ALK} Anderson M H, Ensher J R, Matthews M R, Wieman C E and
Cornell E A 1995 {\it Science} {\bf 269} 198\\
Davis K B, Mewes M O,
Andrews M R, van Druten N J, Durfee D S, Kurn D M and Ketterle W 1995 {\it Phys.
Rev. Lett.} {\bf 75} 3969\\
Bradley C C, Sackett C A, Tollett J J
and Hulet R G 1995 {\it Phys. Rev. Lett.} {\bf 75} 1687

\bibitem{MIT} Killian T C, Fried D G, Willmann L, Landhuis D, Moss S C,
Greytak T J and Kleppner D 1998 {\it Phys. Rev. Lett.} {\bf 81} 3807\\
Fried D G, Killian T C, Willmann L, Landhuis D, Moss S C, Kleppner D and
Greytak T J 1998 {\it Phys. Rev. Lett.} {\bf 81} 3811

\bibitem{HEL} Robert A, Sirjean O, Browaeys A, Poupard J, Nowak S, Boiron D, Westbrook C I,
Aspect A  2001 {\it Science} {\bf 292} 461\\
Pereira Dos Santos F, Leonard J, Wang J, Barrelet C J, Perales F, Rasel E, Unnikrishnan C S,
Leduc M, Cohen-Tannoudji C 2001 {\it Phy. Rev. Lett.} {\bf 86} 3459

\bibitem{RMP} Dalfovo F, Giorgini S, Pitaevskii L P and Stringari S 1999
{\it Rev. Mod. Phys.} {\bf 71} 463 

\bibitem{NEAR} Idziaszek Z, Rza\.{z}ewski K and Lewenstein M 2000
{\it Phys. Rev. A} {\bf 61} 053608


\bibitem{BOX} Hauge E H 1969 {\it Physica Norvegica} {\bf 4} 19\\
Fujiwara I,ter Haar D and Wergeland H 1970 {\it J. Stat. Phys.} {\bf 2} 329

\bibitem{ZIF} Ziff R M, Uhlenbeck G E and Kac M 1977 {\it Phys. Rep.} {\bf 32} 169


\bibitem{BOR} Borrmann P, Harting J, M\"{u}lken O and Hilf E R 1999 {\it Phys. Rev. A} {\bf 60}, 1519 

\bibitem{WIL} Wilkens M and Weiss C 1997 {\it J. Mod. Opt.} {\bf 44} 1801

\bibitem{POL} Politzer H D 1996 {\it Phys. Rev. A} {\bf 54} 5048

\bibitem{GAJ} Gajda M and Rza\.{z}ewski K 1997 {\it Phys. Rev. Lett.} {\bf 78} 2686\\
Navez P, Bitouk D, Gajda M, Idziaszek Z and Rza\.{z}ewski K
1997 {\it Phys. Rev. Lett.} {\bf 79} 1789\\
Holthaus M, Kalinowski E and Kirsten K 1998 {\it Ann. Phys. (N. Y.)} {\bf 270} 198

\bibitem{GRO} Grossmann S, Holthaus M 1997 {\it Phys. Rev. Lett.} {\bf79} 3557

\bibitem{HOL2} Holthaus M, Kalinowski E 1999 {\it Ann. Phys. (N. Y.)} {\bf 276} 321.

\bibitem{GIO} Giorgini S, Pitaevskii L P and Stringari S 1998 {\it Phys. Rev. Lett.} {\bf 80} 5040

\bibitem{IDZ1} Idziaszek Z, Gajda M, Navez P, Wilkens M and Rza\.{z}ewski K 
1999 {\it Phys. Rev. Lett.} {\bf 82} 4376

\bibitem{IDZ2} Illuminati F, Navez P and Wilkens M 1999 {\it J. Phys. B} {\bf 32} L461 

\bibitem{MEI} Meier F and Zwerger W 1999 {\it Phys. Rev. A} {\bf 60} 5133

\bibitem{KOC} Kocharovsky V V, Kocharovsky Vl V and Scully M O
2000 {\it Phys. Rev. Lett.} {\bf 84} 2306

\bibitem{GRA} Graham R 2000 {\it Phys. Rev. A} {\bf 62} 023609

\bibitem{GIO1} Giorgini S, Pitaevskii L P and Stringari S 1996  {\it Phys. Rev. A}
{\bf 54} R4633\\
Naraschewski M, Stamper-Kurn D M 1998 {\it Phys. Rev. A} {\bf 58} 2423\\
Liu S J, Huang G X, Ma L, Zhu S H and Xiong H W 2000 {\it J. Phys. B} {\bf 33} 3911

\bibitem{XIONG} Xiong H W, Liu S J, Huang G X, Xu Z J and Zhang C Y
2001 {\it J. Phys. B} {\bf 34} 3013

\bibitem{DAR} Darwin C G and Fowler R H 1922 {\it Phil. Mag.} {\bf 44} 450, 823;
1922 {\it Proc. Combridge Phil. Soc.} {\bf 21} 262

\bibitem{ROB} Robinson J E 1951 {\it Phys. Rev. } {\bf 83} 678\\
London F 1954 {\it Superfluids} (Wiley, New York) Vol. II Appendix p.203

\bibitem{DEB} Dingle R B 1949 {\it Proc. Cambridge. Phil. Soc} {\bf 45} 275\\
Fraser A R 1951 {\it Phil. Mag.} {\bf 42} 165

\bibitem{PAR} Pathria R K 1972 {\it Statistical Mechanics} (Pergamon Press: New
York) 

\bibitem{HUANG} Huang K 1987  {\it Statistical Mechanics} (John Wiley and Sons, New York)

\bibitem{BOG} Bogoliubov N 1947 {\it J. Phys. USSR} {\bf 11} 23

\bibitem{GIO2} Giorgini S, Pitaevskii L P and Stringari S 1997 {\it J. Low. Temp. Phys.}
{\bf 109} 309

\end{references}
\end{document}